\documentclass[review]{elsarticle}
\usepackage{amsmath,amssymb,latexsym,graphicx,color,hyperref,subfig, graphicx}
\newcommand{\ff}{\mbox{\bf f}}

\newcommand{\Rcal}{\ensuremath{\mathcal{R}}}

\newcommand{\bff}{\ensuremath{\mathbf{f}}}

\newcommand{\bu}{\ensuremath{\mathbf{u}}}

\newcommand{\bw}{\ensuremath{\mathbf{w}}}
\newcommand{\bx}{\ensuremath{\mathbf{x}}}
\newcommand{\by}{\ensuremath{\mathbf{y}}}

\begin{document}

\begin{frontmatter}
\title{Complex dynamics of long, flexible fibers in shear}

\author{John LaGrone}
\ead{lagrone@tulane.edu}
\address{$^1$Department of Mathematics, Tulane University, New Orleans, LA 70118}

\author{Ricardo Cortez}
\ead{rcortez@tulane.edu}
\address{$^1$Department of Mathematics, Tulane University, New Orleans, LA 70118}

\author{Wen Yan}
\ead{wyan@flatironinstitute.org}
\address{$^2$Flatiron Institute, Simons Foundation, New York, NY 10010}

\author{Lisa Fauci}
\ead{fauci@tulane.edu}
\address{$^1$Department of Mathematics, Tulane University, New Orleans, LA 70118}

\date{\today}

\begin{abstract}
The macroscopic properties of polymeric fluids are inherited from the material properties of the fibers embedded in the solvent.
The behavior of such passive fibers in flow has been of interest in  a wide range of systems, including cellular mechanics, 
nutrient aquisition by diatom chains in the ocean,
and industrial applications such as paper manufacturing.  
The rotational dynamics and shape evolution of fibers in shear depends upon the slenderness of the
fiber and the non-dimensional ``elasto-viscous" number that measures the ratio of the fluid's viscous forces to    
the fiber's elastic forces.  For a small elasto-viscous number, the nearly-rigid fiber rotates in the shear, but 
when the elasto-viscous number reaches a threshhold, buckling occurs.  For even larger elasto-viscous numbers, there is a transition to
a ``snaking behavior" where the fiber remains aligned with the shear axis, but its ends curl in, in opposite directions.
These experimentally-observed 
behaviors have recently been characterized computationally
using slender-body theory and immersed boundary computations.  
However, classical experiments with nylon fibers and recent experiments with actin filaments have
demonstrated that for even larger elasto-viscous numbers, multiple buckling sites and coiling can occur.  
Using a regularized Stokeslet framework coupled with a kernel independent fast multipole method, we present simulations that capture these
complex fiber dynamics. 

\end{abstract}

\end{frontmatter}

\section{Introduction}
The motion of flexible fibers in flow is central to many biological systems at the microscale.  
Mammalian sperm flagella propel these cells through the 
female reproductive tract \cite {fd2006}, while microtubule fibers are ingredients of the mitotic spindle in
cell division \cite{shelley2016}.  While the dynamics of these fiber-fluid systems are actuated by molecular motors, other biological  
systems contain passive fibers that are
transported and undergo shape deformations due to the flow.  Examples of these passive fibers include microtubules transported by
cytoplasmic streaming in fungal hyphae \cite {jedd2015} and chains of diatom cells that move with water currents through the ocean \cite {karp1998motion}.     

%The motion of thin flexible fibers in flow also has industrial applications such as paper manufacturing \cite{Stockie98simulatingthe}
Early experiments by Forgacs and Mason \cite{Forgacs1} on synthetic fibers in shear  
demonstrated a spectrum of orbits and shape deformations. 
Shorter, stiffer fibers experienced a signature Jeffery orbit, where periodic tumbling was accompanied by little to no
deformation.  Longer fibers exhibited  periodic orbits with shape deformations that were qualitatively catalogued as S-turns (buckling) and snake turns.  For the
longest fibers, Forgacs and Mason \cite{Forgacs1} observed complex shape deformations that they described as coiled orbits with entanglement. 
Decades later, with the availabilty of microfluidic technology, these 
periodic shape deformations in shear have been observed in DNA strands and actin filaments \cite{Kantsler1,Harasim1,liu2018morphological}. 
Most recently, 
the complex coiling and 
entanglement of long actin fibers was measured in \cite{yananthesis}, where an actin filament of length of more than sixty microns subjected to a shear flow.  A fiber that is initially straight and aligned with the shear direction develops the shape of a hook at the ends of the fiber during a snake turn. Later in time, the fiber exhibits a more complex behavior, including multiple buckling sites in the middle of the fiber and three-dimensional entanglement.
%Figure \ref {yanan} shows snapshots of an actin filament of length of more than sixty microns subjected to a shear flow.            
%While the fiber was initially straight and aligned with the shear direction, the first panel shows the typical hook formations that emerge at the ends of the fiber
%durinig a snake turn.
%The complex behavior is exhibited in the second and third panels that show multiple buckling sites in the middle of the fiber and the 
%three-dimensional entanglement, respectively.

In a review article on the dynamics of flexible fibers in flow,  du Roure et. al. \cite{annrev2019} describe recent technological advances in 
experimentation and recent algorithmic advances in computational modeling that have given rise to deeper understanding and probing  of fiber-fluid systems. 
Exploiting the inertia-free environment at the microscale, and the slender geometry of the fibers, much progress has been made in using 
slender body theory \cite{Pozrikidis:92} to describe the orbits and buckling of fibers in flow e.g. \cite{ts2007, Quennouz1, Manikantan1, liu2018morphological}.
In this manuscript, we present a mathematical model and numerical method that captures the complex shape deformations of the longest fibers, that is
not easily achieved with a slender body formulation.  The fibers we consider need not be thin and are not represented purely by their centerlines, but
by a discretization of their surface.  We use a similar fiber model as that used to examine the dynamics of diatom chains in a non-zero 
Reynolds number environment \cite{Nguyen2017}. 
%coupled to an adaptive-mesh immersed boundary formulation.    
However, here we assume that the length and velocity scales are small enough so that the fluid dynamics is well-described by the Stokes equations,
and use a regularized
Stokeslet formulation \cite{cortez2001, cortez2005method} of the fluid-fiber system. 

While the diatom chain model of \cite{Nguyen2017} was able to capture complex buckling behavior of long fibers, it was based on
an adaptive mesh immersed boundary method that needed fine grid resolution on the fluid domain near the fiber. 
In contrast, the regularized Stokeslet formulation described here, while requiring fine resolution of the fiber surface to capture
the complex behavior, is based on fundamental solutions of the Stokes equations and does not require a spatial grid on the surrounding fluid.   
Although one of the most attractive features of the regularized Stokeslet framework is the ease of implementation -- the velocities at $N$ nodes
are computed based upon the forces at each of the $N$ nodes -  the direct $N^2$ evaluation becomes costly for $N$ large.  
In order to
resolve the surface of a long, thin fiber using a discretization such that the distance between nodes around a cross section is on
the order of the distance between cross sections, 
the number of nodes $N$ becomes necessarily large. 
Here we demonstrate that the incorporation of a kernel independent fast multipole method \cite {ying_kernel-independent_2004} to compute
velocities rather than a direct evaluation allows for faster simulations of the longest fibers, and will be a promising tool for multi-fiber investigations. 

In the following sections we will discuss the construct of the model fiber and its coupling to a Stokes fluid using the method of regularized Stokeslets.  We will demonstrate the shape deformations of fibers of increasing length in shear, and discuss how these results 
compare with recent  studies.  Moreover, we will present simulations  of long fibers (at large elasto-viscous numbers) that capture the
complex coiling and entanglement observed in experiments.

%\begin{figure}
%        \centering
%     \includegraphics[width=0.85\textwidth]{{{yanan1}}}
%     \includegraphics[width=0.85\textwidth]{{{yanan2}}}
%     \includegraphics[width=0.85\textwidth]{{{yanan3}}}
%        \caption{Snapshots of long actin filament under shear, courtesy of Y. Liu \cite{yananthesis}.}
%       \label{yanan}
%\end{figure}

\section{Methodology} \label{methods}

\subsection{Stokes equations}

Assuming that length and time scales are small, we model a flexible fiber coupled to a viscous fluid
in three-space by the 
incompressible Stokes equations:

\begin{align}
	\begin{split}
	0 &= -\nabla \hat{P} + \mu \Delta \hat{\mathbf{u}} + \hat{\mathbf{F}} , \\
	0 &= \nabla \cdot \hat{\mathbf{u}},
\label{stokes}
	\end{split}
	\end{align}
	where $\hat{P}$ is the pressure, $\hat{\mathbf{u}}$ is the fluid velocity, $\mu$ is the fluid viscosity, 
	 and  $\hat{\mathbf{F}}$ is the external force per  
volume exerted by the fiber on the fluid.
%	Variables with a ``hat" are dimensional; without it are dimensionless.
The forces in  
Eqn. \ref{stokes} are localized at the fiber surface, and will be described below.   
The motion of the passive fiber will be driven by an imposed background shear flow ${\hat{\bf u}}_b (\hat{x},\hat{y},\hat{z})  = (\dot {\gamma} \hat {y}, 0, 0)$. 

%0 = - \nabla p +\Delta\bu 
%+   \FF 
%\hskip20pt\mbox{and}\hskip20pt
%\frac{\hat\ell}{T} \bu = \frac{{\mathcal F}}{\mu\hat\ell }S[\FF] 
%+ \frac{{\mathcal L}}{\mu } R[{\bf L}] ,  
%\]
\noindent
%Throughout this manuscript, we choose the characteristic scales to be: 
%$\hat\ell = $ 
%4 $\mu$m,  
%$\hat{T} = $
%.01 sec,  
%$\mu$ to be the viscosity of water, and, hence, $\cal F$ =  $1.6\times10^{-12}$ N.

We  use a regularized Stokeslet framework \cite{cortez2005method} to model the elastohydrodynamic system.
Rather than using a surface integral of Dirac delta function forces $A(\by) \hat{\ff}(\by) \delta(\bx-\by)$, we consider
regularized force density  $A(\by)\hat{\ff}(\by) \phi_{\varepsilon}(\bx-\by)$ supported on a patch of area  $A(\by)$ on the surface of the fiber.
%The force is:    
%
%\[
%\FF(\bx) = \int_\Sigma \ff(\by) \phi_{\varepsilon}(\bx-\by) dS_y
%%,\ \ \ 
%%\mathbf{L}(\bx) = \sum_{k=1}^2 \bg_k \phi_{\varepsilon}(\bx-\by_k).
%\]
The regularization (or blob) function is chosen to be:
        \begin{align}
        \phi_\varepsilon (\mathbf{x} - \mathbf{y}) = \frac{15 \varepsilon^4}{8 \pi (r^2 + \varepsilon^2)^{7/2}},
        \end{align}
        where $r = \| \mathbf{x} - \mathbf{y} \|$, and $\varepsilon$ is the regularizaiton parameter \cite{cortez2005method}.
This leads to the velocities due to the  regularized Stokeslets as follows: 
	\begin{align} \label{eq:stokelet}
{\mathbf{u}}_{st}(\mathbf{x}) = \int_\Sigma S_\varepsilon(\mathbf{x},\mathbf{y}) {\mathbf{\hat{f}}}(\by) dS_y
	= \frac{1}{8 \pi \mu} \int_\Sigma \frac{(r^2 + 2 \varepsilon^2) {\mathbf{\hat{f}}}(\by)  
+ ({\mathbf{\hat{f}}}(\by) \cdot (\mathbf{x} - \mathbf{y})) (\mathbf{x} - \mathbf{y})}{(r^2 + \varepsilon^2)^{3/2}} dS_y, 
	\end{align}

%    \begin{align} \label{eq:rotlet}
%    {\mathbf{u}}_{rt}(\mathbf{x}) = \sum_{k=1}^{2} R_\varepsilon(\mathbf{x},\mathbf{y}_k) \mathbf{g}_k = \frac{1}{16 \pi} \sum_{k=1}^{2} \frac{2r_k^2 + 5 \varepsilon^2}{(r_k^2 + \varepsilon^2)^{5/2}} \left(\mathbf{g}_k \times (\mathbf{x} - \mathbf{y}_k) \right),
%    \end{align}
where ${\mathbf{\hat{f}}}(\by)$ is force per unit area and  $\Sigma$ denotes the surface of the fiber.
As the regularization parameter $\varepsilon$ approaches zero, the kernel $S_{\varepsilon}$ approaches the classical singular
Stokeslet $S_{0}$.

We nondimensionalize this coupled fluid-fiber  problem by  choosing  a characteristic length scale
$\mathcal{L}=\hat{L}_{0}$ on the order of a fiber length,  
a time scale $\mathcal{T}=\alpha \dot{\gamma}^{-1}$,
a velocity scale $\mathcal{U}= \hat{L}_{0}/\mathcal{T}$,  
and
a force scale  ${{\mathcal F}}
=\mu \hat{L}_{0}^2/\mathcal{T}$.   
Here $\alpha$ is a non-dimensional tuning parameter for background shear flow that is chosen to be 
$\alpha = 2.5$ in all simulations shown below.
We will use these non-dimensionalized quantities throughout the
manuscript.

%%%%%%%%%
\subsection{Representation of fiber}
%%%%%%%%%

\begin{figure}
                \centering
                \includegraphics[width=\textwidth]{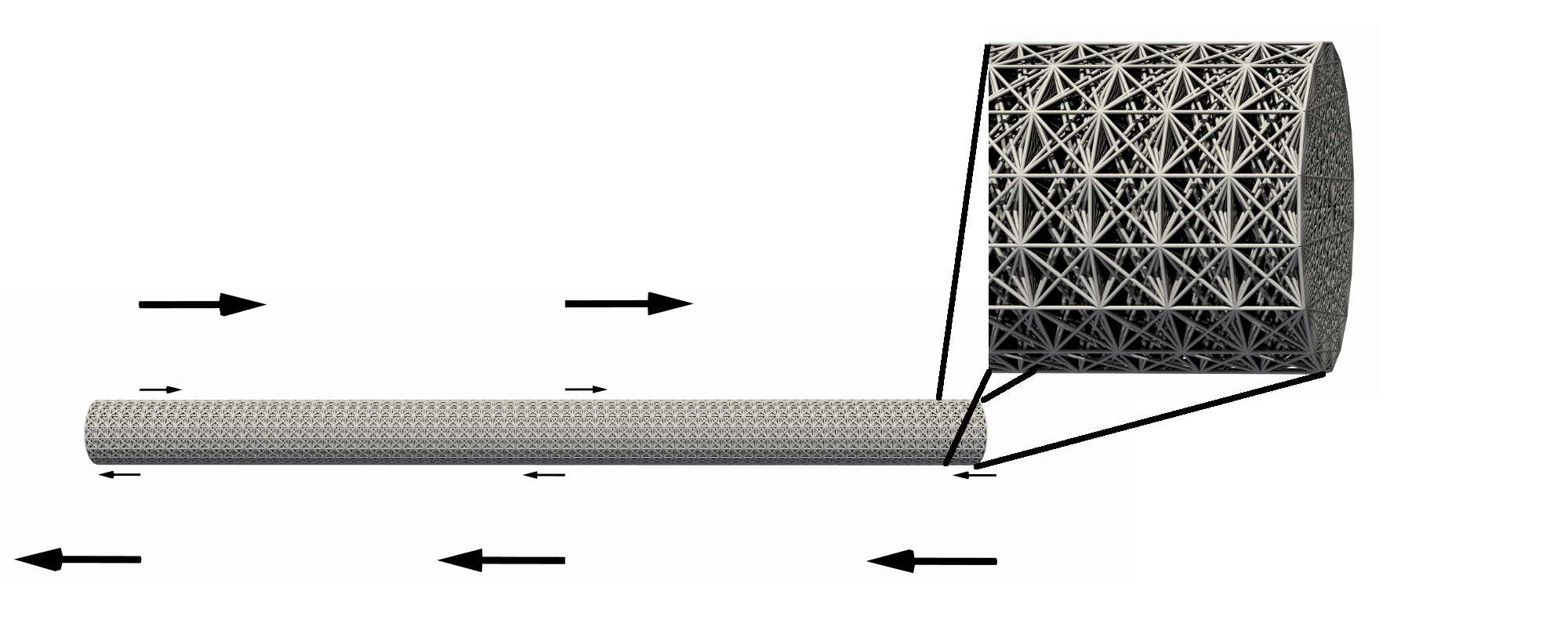}
                \caption{Fiber model consisting of points on fiber surface  connected by a network of springs.  The inset shows
more detail of the spring network comprising  the fiber. 
The arrows indicate the background shear flow.} 
       % \end{subfigure}
       \label{schematic}
        \end{figure}

The model fiber that we consider has a native straight shape and equilibrium length $L$. 
We construct the discretization of the surface of the cylindrical fiber by 
placing cross-sections of radius $R_f$ along the centerline, perpendicular to the centerline
(see Figure \ref{schematic}). 
Each cross-section is discretized by $N_c$ = 18  points, and we take $N_f$ cross-sections along the fiber 
so that the spacing between neighboring cross-sections is approximately
equal to the spacing between adjacent points on a cross section.
% (see Figure \ref {fig:flagella}).  

Each of the $N = N_c \times N_f$ discrete  points on the surface of the fiber is connected to a subset of
the other surface points by a Hookean spring,  giving elasticity to the structure.  We define the 
elastic energy in the system as
	\begin{align}\label{eq:Energy}
{\cal E} = \frac{1}{2} \sum_j k_j  l_j \left(\frac{\| \bx_{j_1}-\bx_{j_2} \|}{l_j}-1 \right)^2 
%= \frac{1}{2} (EI) \beta^2 L
	\end{align}
where $k_j$ is the stiffness of a spring with resting length $l_j$ that connects points
$j_1$ and $j_2$.  The sum is over all springs.
The force at $\bx_{j_1}$ is ${\ff}_{j_1}A_{j_1}$ where $A_{j_1}$ is  the area of a patch of surface
centered at $\bx_{j_1}$ in the discretization.  The elastic forces are derived from the energy function:
\[
{\ff}_{j_1}A_{j_1} = - \frac{\partial {\cal E}}{\partial \bx_{j_1} }   
\]

This network of nodes and elastic linkages will impart tensile stiffness and bending rigidity to the fiber, calibrated by
the connectivity of the nodes and the stiffness constants of individual linkages. 
Similar constructs of semi-flexible filaments coupled to an incompressible fluid have been 
used to model
bacterial flagella \cite {lagrone1,peskinlim2004,Flores2005}
and diatom chains \cite{Fauci4}. 
In all simulations shown here, we choose a network connectivity 
so that each point on a given cross-section is connected to every other point on that cross-section, as well as to every other
point on the two cross-sections adjacent to it.  This means that each node is connected to $17 +  2 \times 18 = 53$ other nodes.  
In addition, in all simulations shown, the
stiffness constant $k_j = k$ in Eqn. \ref{eq:Energy}  is taken to be the same for all springs.  
The resting lengths of the springs, $l_j$ in Eqn. \ref {eq:Energy} do vary with $j$, and are determined
by the straight fiber configuration.

\begin{table}
                 \label{bigtable}
                \begin{center}
                        \begin{tabular}{|l|c|}
                                \hline
                                Quantities & Dimensionless  \\
                                      &  value      \\
                                \hline
                                Fiber length, $L$                       & $0.139 - 2.28$         \\
                                Fiber radius, $R_f$             & $0.005$  \\
                                Slenderness ratio, $\rho = R_f/L$             & $0.0022 - 0.036$    \\
                                Spring stiffness, $k$              & $0.0112$    \\
                                Bending rigidity, $EI$                                   & $4.8\times 10^{-6}$ \\% $4.3 \times 10^{-9}$   \\
                                Shear scale,  $\alpha $              & $2.5$    \\
                                Elasto-viscous number, $\bar{\mu}$                                   & $5.35\times 10^{2} - 2.37 \times 10^{7}$   \\
%                                Spring Energy, $E$ & 0.013 - 14.8 & $8.5 \times 10^{-20} - 9.5 \times 10^{-17}$ J \\
                                \hline
                                Numerical Parameters   &  \\
                                \hline
                                Cross sections along fiber, $N_f$                       & $80 - 1313 $         \\
                                Points per fiber cross section, $N_c$                   & $18$        \\
                                Total number of points on fiber $N $                   & $1440 - 23634$        \\
                                Spacing between fiber nodes, $\Delta s$ & $ 0.00174$           \\
                                Blob size, $\varepsilon$ &0.00225  \\
                                Time step, $\Delta t$   & $1.0 \times 10^{-4} $    \\ 
                                \hline
                        \end{tabular}
\caption{Geometric and material parameters of fibers and numerical parameters (all non-dimensional).}
                \end{center}
\end{table}

Due to the imposed background shear, the flexible fiber will depart from 
its equilibrium shape as the network springs become stretched or compressed, 
causing forces at the nodes to develop.  
The fluid velocity due to these elastic forces 
is evaluated at each material point $\mathbf{x}_i, i = 1,...,N$ 
of the fiber surface, using 
a discrete version of 
(Eq.~\eqref{eq:stokelet}): 

\[
\bu_{st}(\mathbf{x}_i) = 
\sum_{j=1}^{N} S_\varepsilon(\mathbf{x}_i,\mathbf{x}_j) {\ff}_{j}A_{j}.
\]

Using the forward Euler method, this velocity, added to the background shear velocity $\bu_{b}$,  
is used to update the positions of the nodes of the flexible fiber.  
The range of the fiber's (non-dimensional) geometrical and elastic parameters used in simulations are shown in Table 1,
along with the numerical parameters used.

	\subsection{Calculation of fiber bending rigidity  and non-dimensionalization} \label{sec:sperm_number}

The macroscopic bending rigidity $EI$ of the node-spring structure depends upon the
individual spring constants $k _ j $ and  the topology of the spring network.  
Following \cite{peskinlim2004, Fauci4}, we precompute the bending rigidity $EI$ for the
model fiber by  bending it into a 
circular arc with a prescribed radius of curvature $\kappa$.  Because this curved shape stretches and compresses the
network of springs, a non-zero elastic energy $\mathcal {E}$ 
from  Eqn. \ref{eq:Energy} results, arriving at: 

	\begin{align} 
%	\xi^\perp &= \frac{4  \pi }{\log \left(\frac{L}{a}\right) + 1},  \label{eq:xi_perp}\\
	{EI} &= \frac{2{\cal E}}{\kappa^2 L}.
	\end{align}

As in other elastohydrodynamic systems where flexible fibers are coupled to a Stokesian incompressible fluid, 
the dynamics are governed by two non-dimensional parameters: the fiber aspect ratio (slenderness parameter)  $\rho = R_f/L$, and the elasto-viscous number that
measures the 
relative importance of flow forces to
elastic forces (e.g. \cite{ts2007,Fauci4, qiang2017, Wandersman1, liu2018morphological}).
We define the elasto-viscous number:

%	\begin{align}
%        \label{spermnumber}
%	\end{align}
%	where
%	\begin{align} 
%	\xi^\perp &= \frac{4  \pi }{\log \left(\frac{L}{R_f}\right) + 1}.  \label{eq:xi_perp}\\
%%	\mathbf{EI} &= \frac{2{\cal E}}{\kappa^2 L}.
%	\end{align}

\[
\bar\mu = 
 \frac{8\pi  \mu  \dot\gamma {\hat {L}}^4 }{c \hat{E}\hat{I}} 
%\log(e L^2/R^2)}
=  \frac{8 \pi \alpha  L^4 }{c EI},
%= \frac{4\pi\beta^2\ell^5}{\kappa {\cal L}_0 \rho^2 \log(e \ell^2/\rho^2)}\ \mbox{ which equals }
\]

where we have indicated the ratio both in dimensional parameters and using our non-dimensional scaling. 
For a given slenderness ratio $\rho$, the tangential drag on the filament is given by the geomentric parameter
$c =  \ln ({4 e /{\rho}^2})$.
Moreover, in this work, we do not consider the effect of thermal fluctuations.

\section{Using mixed kernels in Kernel Independent FMM for regularized Stokeslet}

\subsection{Kernel Independent FMM steps}

During the solution of the fiber dynamics, the velocity of each  node on the fiber %Eq.~\eqref{eq:kernelsum} 
must be evaluated at every time step. The direct evaluation of the equation
\begin{align}
\label{eq:kernelsum}
\mathbf{u}(\bx_i) & %= \sum_{j=1}^{N} S_\varepsilon(\mathbf{x}_i,\mathbf{x}_j) \bff_{j}A_{j}.
=\sum_{j=1}^{N} S_\varepsilon(\mathbf{x}_i,\mathbf{x}_j) \bw_j.
\end{align}
requires $O(N^2)$ operations.
The cost can be reduced to $O(N)$ operations with the Kernel Independent Fast Multipole Method (KIFMM) \cite{ying_kernel-independent_2004}, which builds an adaptive octree for a given set of source (S) points $\bx_j$ and target (T) points $\bx_k, 1 \leq k \leq N$, by recursively refining the octree with no more than $N_{leaf}$ points in each leaf box. 

The interactions between points in a leaf box and other points are divided into near-field and far-field.
Near-field interactions refer to points in all adjacent leaf boxes and are summed directly. 
Far-field interactions refer to points in non-adjacent leaf boxes and are approximated using a set of equivalent points with source strength $\bw_k^{equiv}$ on each equivalent point $k$. 
In three dimensions, the equivalent points are chosen to be on a uniform cubic surface mesh surrounding each octree box, with $p$ points along each cubic edge.
The total number of equivalent points is $6(p-1)^2+2$.
This approximation converges exponentially with increasing number $p$.
The contribution from near- and far-fields are added together to form $\bu$ at each target point.

The tradeoff between cost and accuracy is controlled by $p$. In general, $p=10$ gives single-precision accuracy and $p=16$ gives double precision accuracy.
$N_{leaf}$ controls the depth of the octree, and thus affects the total computation time. In practice,
$N_{leaf}$ is set to about 2000 to fit the current CPU architecture.

There are two ways to invoke the KIFMM for $S_\varepsilon$. 
The first choice is to directly use $S_\varepsilon \in \Rcal^{3\times3}$ as the kernel for all operations throughout the octree, for both near-field and far-field interactions. 
This approach is straightforward to implement, but is limited to a common value of $\varepsilon$ for all points.
Further, if periodic boundary condition are necessary, the periodizing operator $\mathcal{T}_{M2L}$ must be recalculated for each different $\varepsilon$~\cite{yan_flexibly_2018}.
Another approach is to regard each source point in $S_\varepsilon$ as a four dimensional vector $(w_1,w_2,w_3,\varepsilon)$.
In this case, $S_\varepsilon$ becomes a nonlinear kernel because $\varepsilon$ appears nonlinearly in $S_\varepsilon$.
The near-field interactions  between source-target pairs are computed directly.
For far-field interactions, the singular kernel $S_0 $ is used for equivalent points, and the strength $\bw_k^{equiv}$ is found by matching the equivalent flow field with the regularized flow field generated by source points.
It is straightforward to find the equivalent $\bw_k^{equiv}$ and the exponential convergence of KIFMM is maintained because the regularized kernel $S_\varepsilon$ is a solution to the Stokes equation.
After $\bw_k^{equiv}$ is found, the singular kernel $S_0$ is used throughout the tree traversal.
This approach allows the regularization parameter $\varepsilon$ to vary for different source points, and allows the direct reuse of $\mathcal{T}_{M2L}$ computed for $S_0$ \cite{yan_flexibly_2018} to implement various types of periodic boundary conditions.

The computation code is implemented based on the parallel KIFMM library \texttt{PVFMM} \cite{malhotra_pvfmm_2015}. 
Both \texttt{MPI} and \texttt{OpenMP} parallelism are implemented to improve parallelization efficiency.
\texttt{SIMD} instructions are also utilized to improve efficiency on modern CPU architectures.

\begin{table}[h]
	\centering
	\begin{tabular}{l | l } 
		\hline
		$S_\varepsilon$ & regularized Stokeslet kernel  \\
		$S_0$ & singular Stokeslet kernel \\
		$\bw_j = \bff_jA_j$ & the source strength at source point $j$ \\
		$\bw_k^{equiv}$ & the equivalent source strength at equivalent point $k$ \\
		$p$ & the number of equivalent points along each cubic box edge \\
		$N_{leaf}$ & the maximum number of points per leaf box \\
		\hline
	\end{tabular}
	\caption{Nomenclature for KIFMM to compute Eq.~\eqref{eq:kernelsum}.}
	\label{tab:Nomenclature}
\end{table}

\section{Results and Discussion}

In recent laboratory experiments that investigate 
the dynamics  of actin filaments in shear \cite{liu2018morphological}, the diameter and bending rigidity of the fibers 
are fixed by nature.  The elasto-viscous number of 
the experiments is varied by either adjusting the background shear or observing actin filaments of different lengths.
Motivated by these experiments, 
here we choose to examine the dynamics of fibers in shear by keeping their bending rigidity and their diameter fixed, but vary their length.
The shear rate remains fixed in these simulations. 
As the fiber length increases, its slenderness ratio decreases, and the corresponding elasto-viscous number of the system increases. 

Although thermal fluctuations play a role in the dynamics of fiber-flow systems, 
%such as the actin fibers in Figure \ref{yanan}
we choose not to include any Brownian forces in the simulations presented here. 
Moreover, 
because we track the surface of the fiber rather than just its centerline as in slender body models, a straight 
fiber that is initialized with its centerline along the $x$-axis needs no perturbation 
to begin its orbit -- spatial gradients in velocity on each cross section are immediately formed.  In the absence of Brownian fluctuations, the dynamics of a 
perfectly cylindrical fiber whose centerline initially coincided with
the $x$-axis in this unbounded shear flow must necessarily obey some symmetry constraints.  In particular, the centerline must remain in the plane that it was initialized in  and the
shape deformations with respect to the centroid of the fiber must be odd. 
In the simulations presented below, we will see the effects of the small fluctuations that occur due to the 
numerical perturbations that arise from time integration and from the 
finite discretization of the fiber surface.% (it is not a perfect cylinder). 

\begin{figure}[t]
                \begin{tabular}{|cc|cc|cc|}
                        \hline
                        \multicolumn{2}{|c|}{$ \bar{\mu} = 5.35\times 10^{2} $} & \multicolumn{2}{|c|}{$ \bar{\mu} = 1.97\times 10^{3}$} & \multicolumn{2}{|c|}{$\bar{\mu} = 8.22\times 10^{4}$} \\
                        \hline
                        T = 0 & \includegraphics[height=2cm]{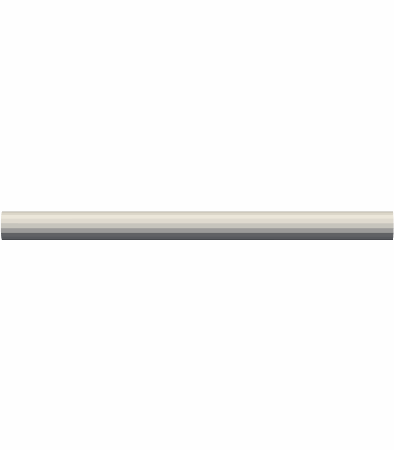}  & T = 0 & \includegraphics[height=2cm]{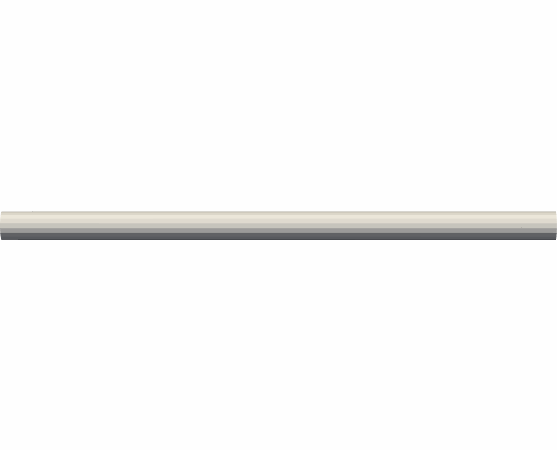} & T = 0 & \includegraphics[height=2cm]{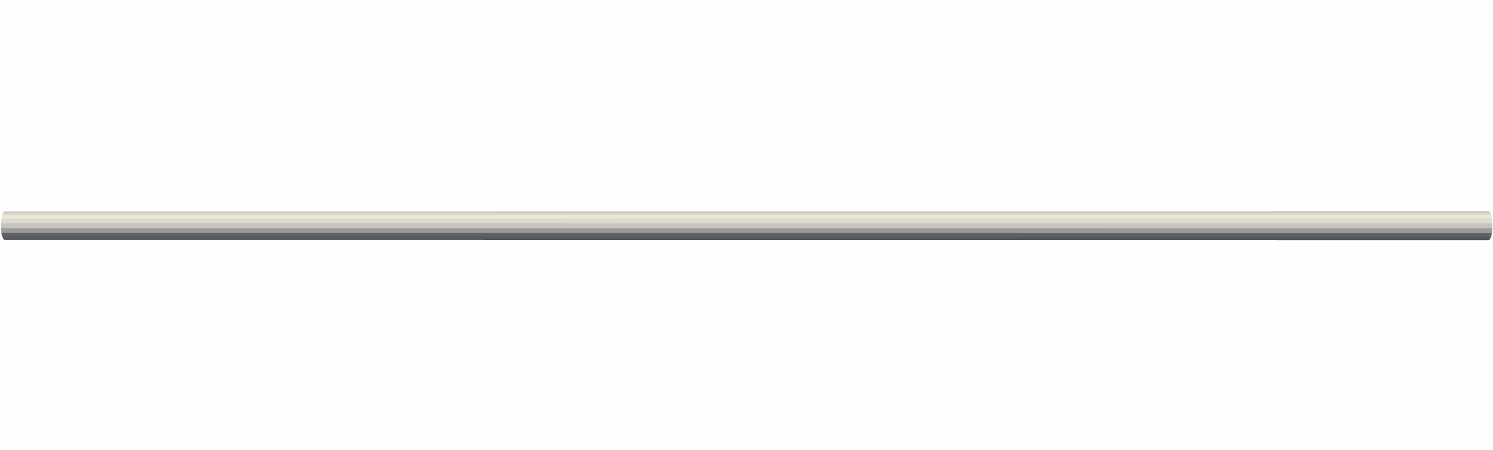} \\
                        \hline
                        T = 4.95 & \includegraphics[height=2cm]{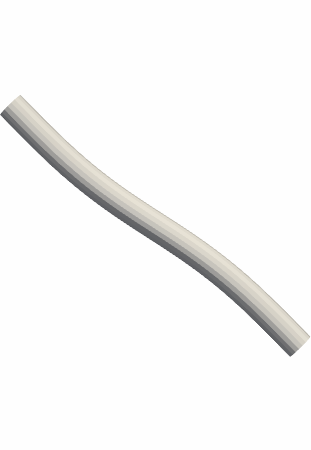}  & T = 5.65 & \includegraphics[height=2cm]{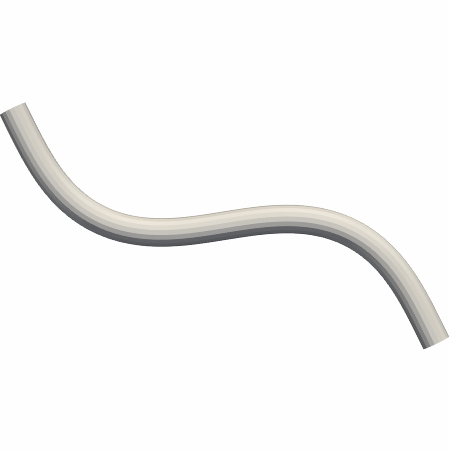} & T = 6.25 & \includegraphics[height=2cm]{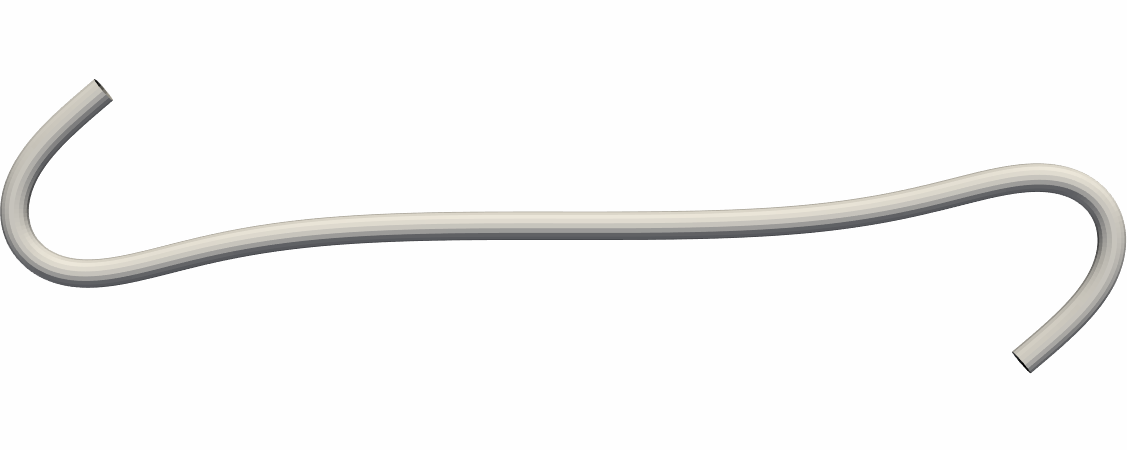} \\
                        \hline
                        T = 5.45 & \includegraphics[height=2cm]{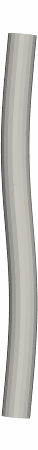}  & T = 6.5 & \includegraphics[height=2cm]{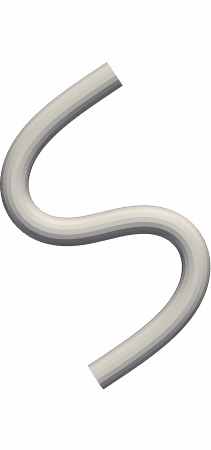} & T = 8 & \includegraphics[height=2cm]{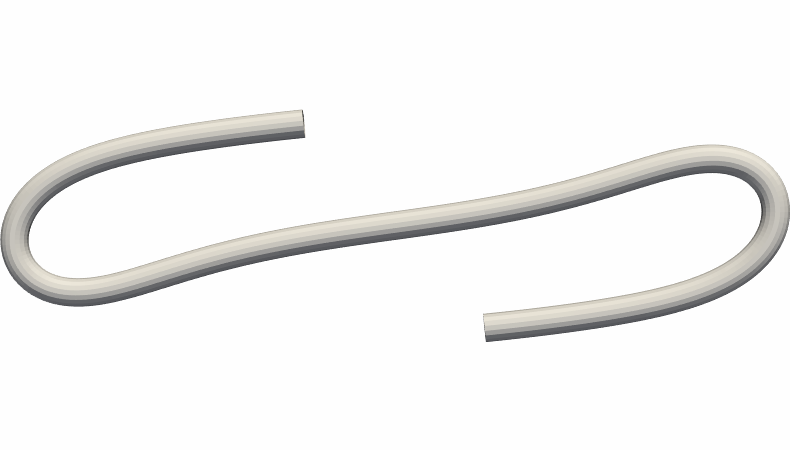} \\
                        \hline
                        T = 5.7 & \includegraphics[height=2cm]{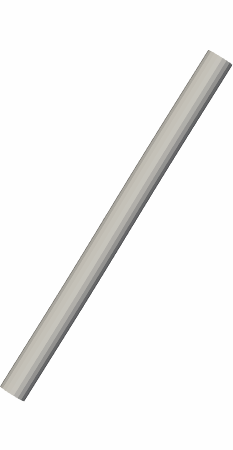}  & T = 7.15 & \includegraphics[height=2cm]{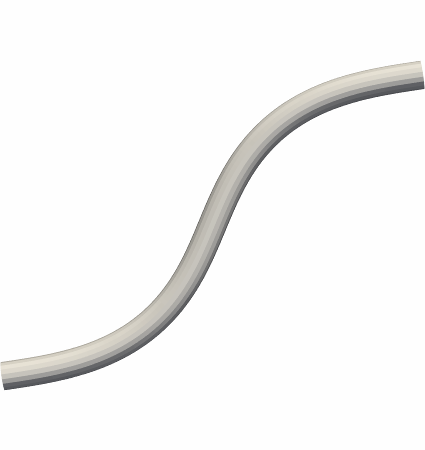} & T = 10.5 & \includegraphics[height=2cm]{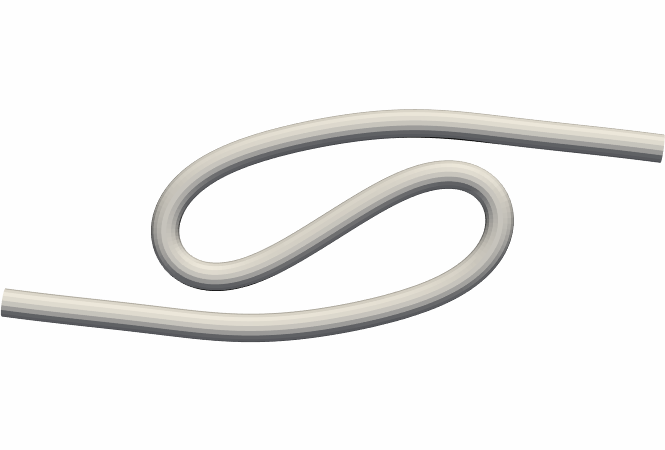} \\
                        \hline
                        T = 7 & \includegraphics[height=2cm]{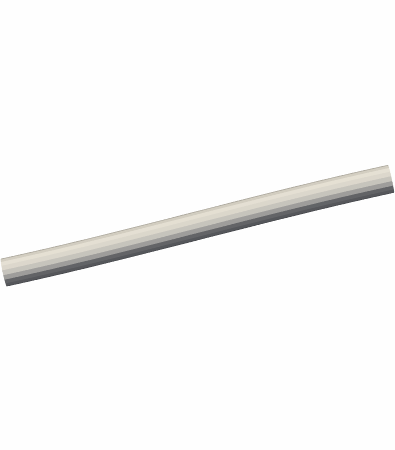}  & T = 8.75 & \includegraphics[height=2cm]{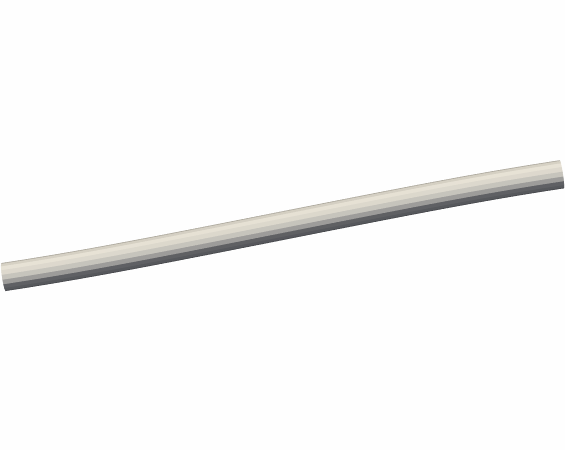} & T = 12.75 & \includegraphics[height=2cm]{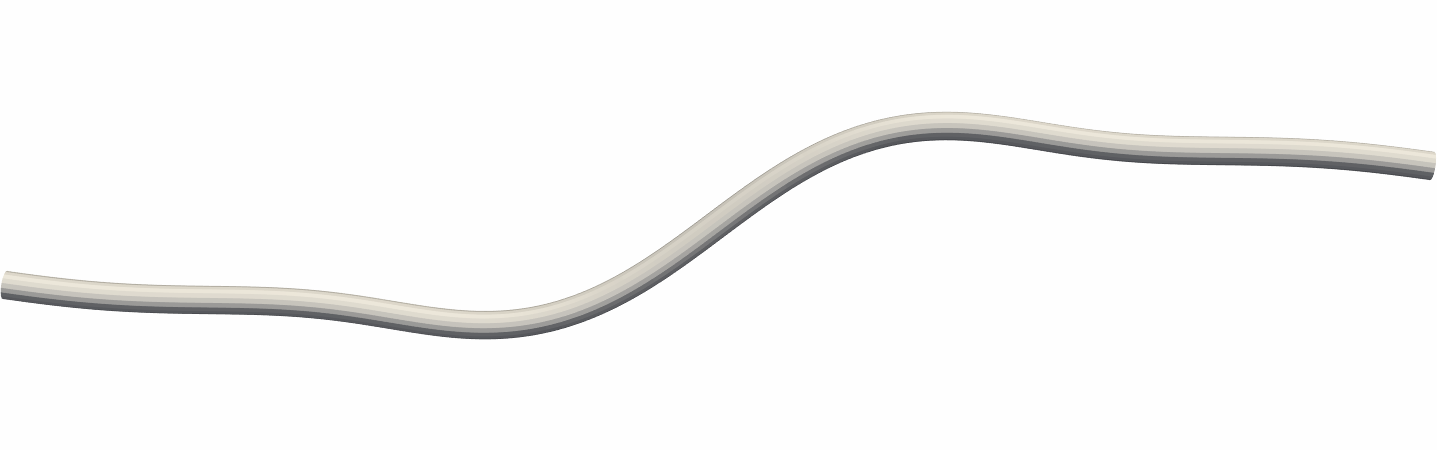} \\
                        \hline
                        T = 10.75 & \includegraphics[height=2cm]{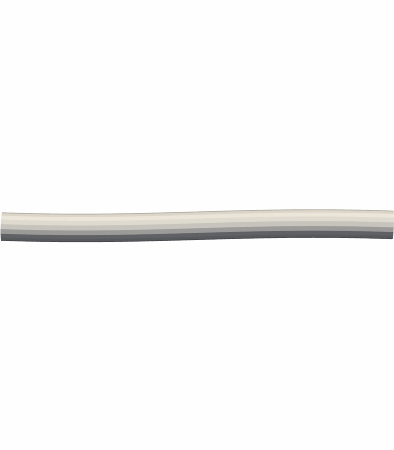}  & T = 13.5 & \includegraphics[height=2cm]{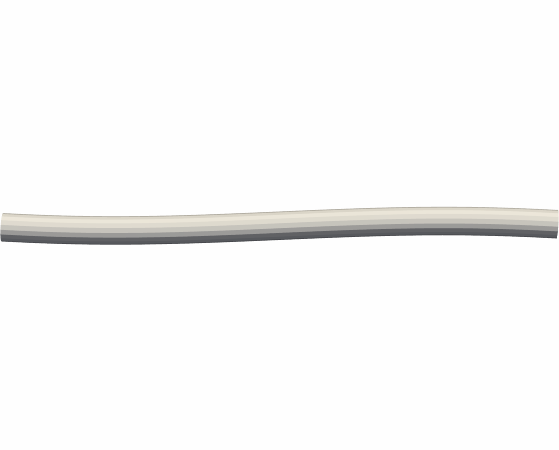} & T = 25 & \includegraphics[height=2cm]{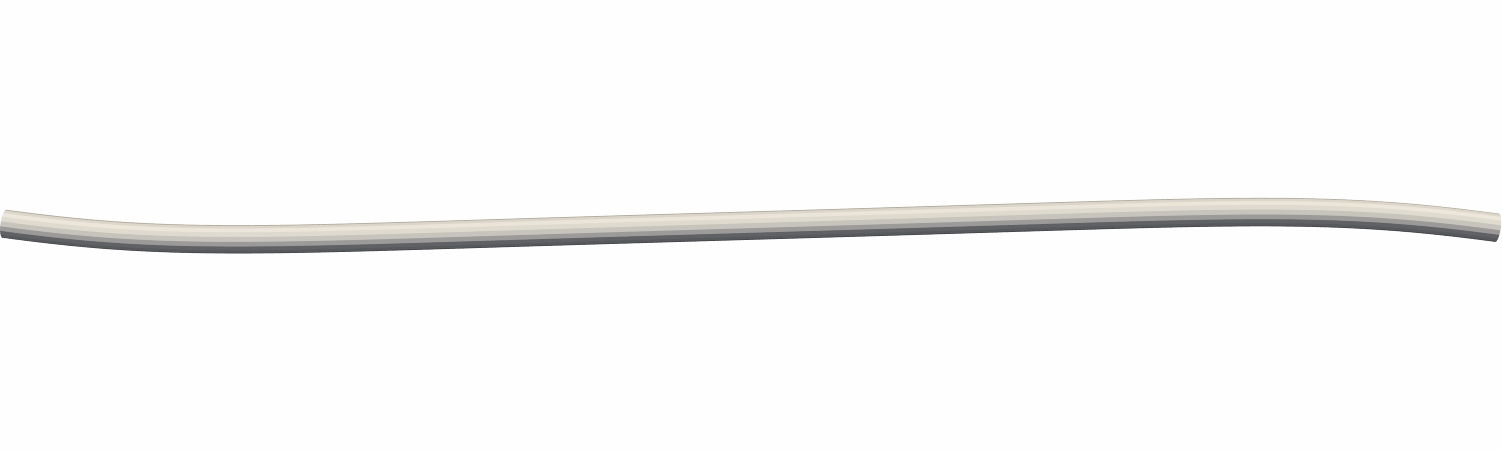} \\
                        \hline
                \end{tabular}
            \caption{Fibers of lengths $L = 0.139, 0.196, 0.523$  with corresponding elasto-viscous numbers of
$\bar{\mu} =  5.35\times 10^{2},  1.97\times 10^{3}, 8.22\times 10^{4}$ respectively. Here we see the classical
tumble, $S$-turn, and snaking periodic orbits. }
           \label{tumble}
        \end{figure}

\begin{figure}
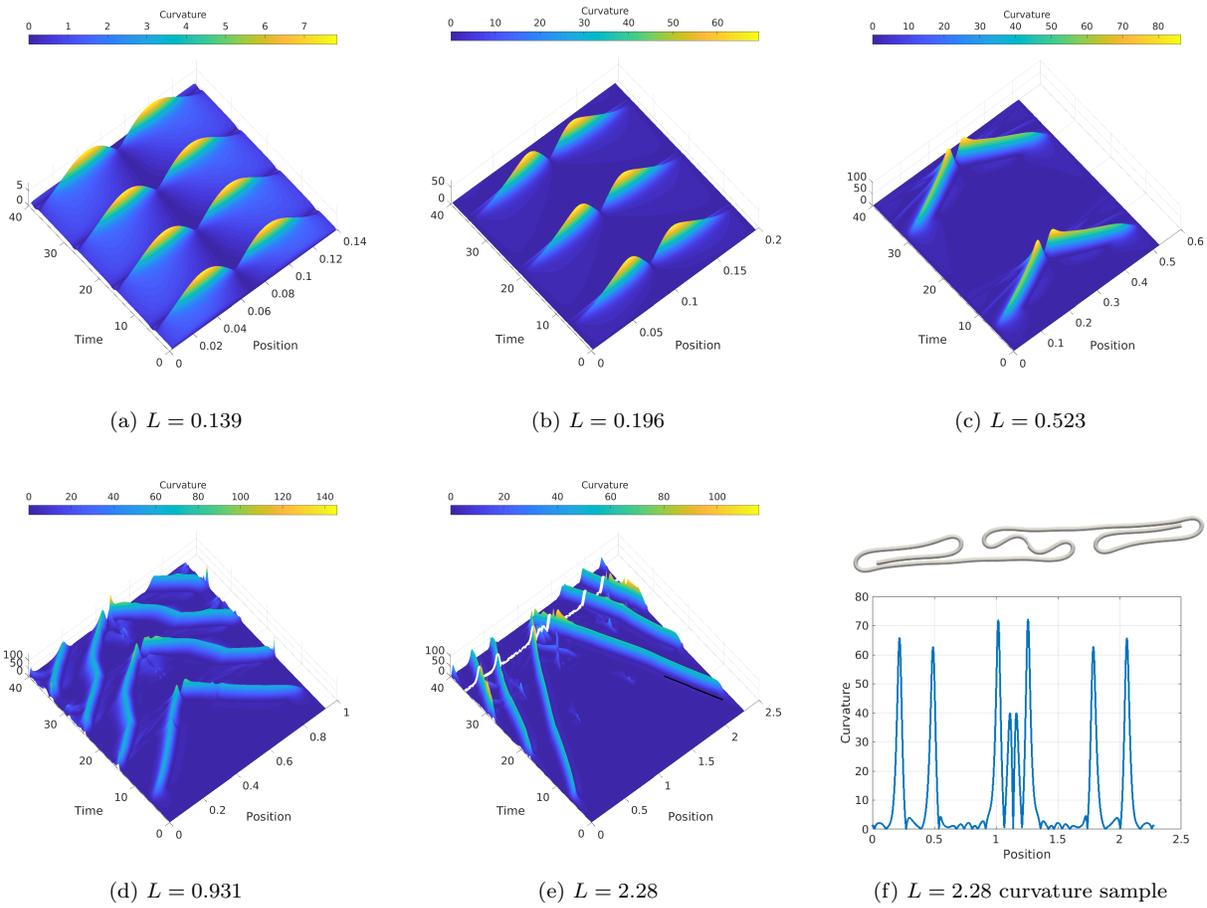

        \centering
        \subfloat[$L=0.139$]{\includegraphics[width=0.32\textwidth]{{{curvature_surface_view2_L=0.138919}}}}\hfill
        \subfloat[$L=0.196$]{\includegraphics[width=0.32\textwidth]{{{curvature_surface_view2_L=0.196222}}}}\hfill
        \subfloat[$L=0.523$]{\includegraphics[width=0.32\textwidth]{{{curvature_surface_view2_L=0.522681}}}}\hfill

        \subfloat[$L=0.931$]{\includegraphics[width=0.32\textwidth]{{{curvature_surface_view2_L=0.930754}}}}\hfill
        \subfloat[$L=2.28$]{\includegraphics[width=0.32\textwidth]{{{curvature_surface_view2_L=2.28_with_line}}}}
         \put(-50,70){\line(5,-2){22}}         \hfill
        \subfloat[$L=2.28$ curvature sample]{\includegraphics[width=0.32\textwidth]{{{curvature_with_fiber}}}}\hfill

        \caption{Surface plots of the curvature as a function of time and arclength along the fiber. Note that
we are not considering signed curvature for fibers of length (a) $L = 0.139$; (b) $L = 0.196$, (c) $L = 0.523$,
(d)$L = 0.931$, (e)$L = 2.28$.  Shown in panel (f) is the curvature along the longest fiber $L = 2.28$ at time $T= 36$.
Here multiple extrema in curvature appear and can be identified with the fibers shape also shown.}  
\label{curvature}
\end{figure}

Figure \ref {tumble} shows the periodic orbits of three fibers of increasing length that display the signature tumble, $S$-turn and snaking behavior reported by \cite {Forgacs1,
liu2018morphological, ts2007, Nguyen2017}. 
The first column of Figure \ref {tumble} depicts the rotational orbit of a fiber of length $L = 0.139\ (\bar{\mu} =
5.35\times 10^{2} $)
 that exhibits 
little deformation from its straight shape.  
Figure \ref{curvature} (a) shows a surface plot of the evolving (absolute value of) curvature
along the arclength of the fiber as a function of time during approximately five orbits.  
We see that two slight bends occur 
during each rotation when the
fiber is aligned with the maximal compressive region of the shear (a forty-five degree angle with the negative $x$-axis),
as in the $T = 4.95$ snapshot in the first column of Figure \ref {tumble}.  The positions of maximal curvature
do not travel along the arclength of the fiber, but occur as standing waves that appear periodically in each orbit.
The second column of Figure \ref {tumble} depicts the rotational orbit of a fiber of length $L = 0.196\ (\bar{\mu} =
 1.97\times 10^{3} $)
that clearly buckles into an $S$-shape during its orbit. 
Figure \ref{curvature} (b) shows the corresponding evolution of curvature.  Again we see standing waves of curvature.
In contrast, the third column of Figure \ref{tumble} 
shows the snaking behavior of a fiber of length $L = 0.523\  (\bar{\mu} =  8.22\times 10^{4})$.  
Here the ends of the fiber  curl in towards the middle of the fiber in an antisymmetric manner, and the points of
maximal curvature travel along its arclength.  These traveling waves of maximal curvature are evident 
in Figure \ref{curvature} (c) during the times at which the fiber is bent from its straight configuration.
In each of the simulations shown in Figure \ref {tumble}, the fibers exhibit periodic orbits, and during most
of the orbit the fibers remain in a nearly straight configuration aligned with the $x$-axis.   
Note that the period of rotation increases with the length of the fiber.

We compare the above with the coordinated experiments and simulations of actin fibers with Brownian fluctuations
by Liu et. al. \cite{liu2018morphological} where the
values of elasto-viscous number $\bar{\mu}$ at which transitions from tumbling to buckling to snaking occurred were
quantified.
The elasto-viscous numbers of the simulations
presented in Figure \ref {tumble} fall squarely within the ranges of these different dynamical regimes.
We note that because of the imposed symmetry in our model, in the absence of a perturbed initial position or Brownian fluctuations, we do not observe the $C$-buckling or $U$-turns reported in \cite{liu2018morphological}, but
rather their counterparts with odd symmetry.  If perturbations were added to the initial placement of the fiber, we do observe such asymmetric shapes (not shown).

\begin{figure}
        \centering
        \begin{tabular}{cc}
                T = 0 & \includegraphics[width=0.8\textwidth]{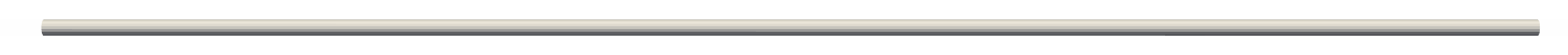} \\
                T = 4.95 & \includegraphics[width=0.8\textwidth]{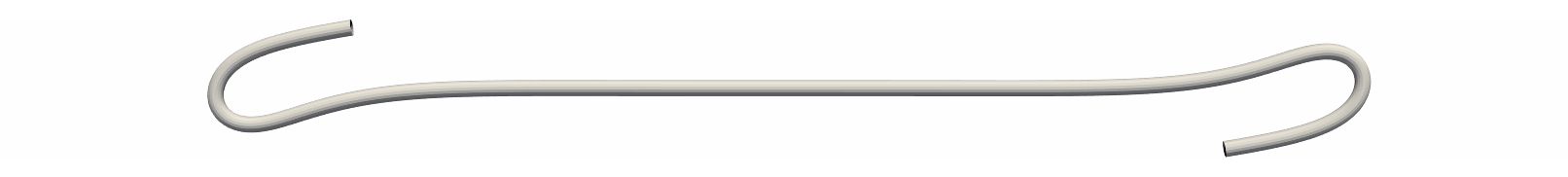} \\
                T = 13.35 & \includegraphics[width=0.8\textwidth]{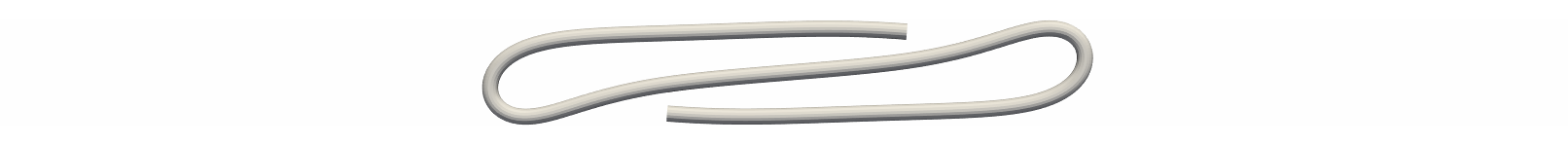} \\
                T = 16.25 & \includegraphics[width=0.8\textwidth]{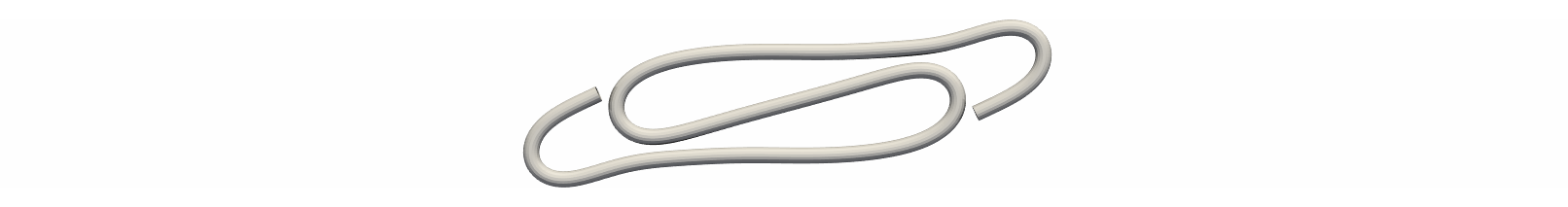} \\
                T = 18.25 & \includegraphics[width=0.8\textwidth]{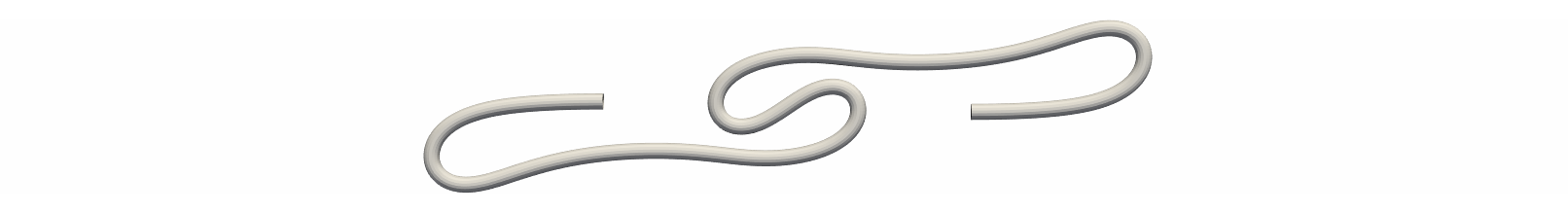} \\
                T = 20.3 & \includegraphics[width=0.8\textwidth]{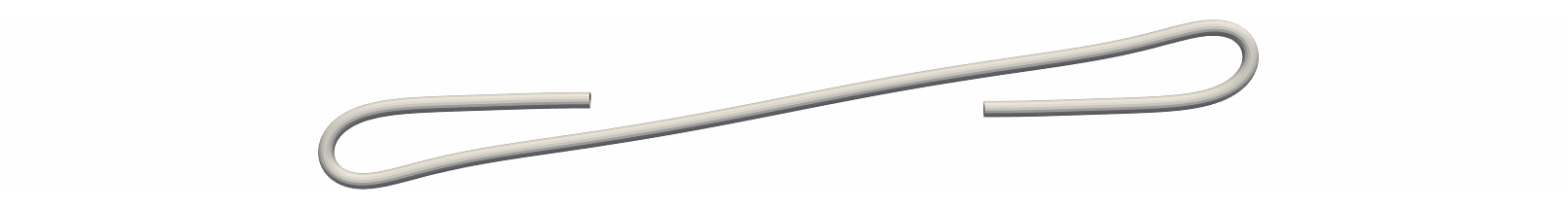} \\
                T = 23.75 & \includegraphics[width=0.8\textwidth]{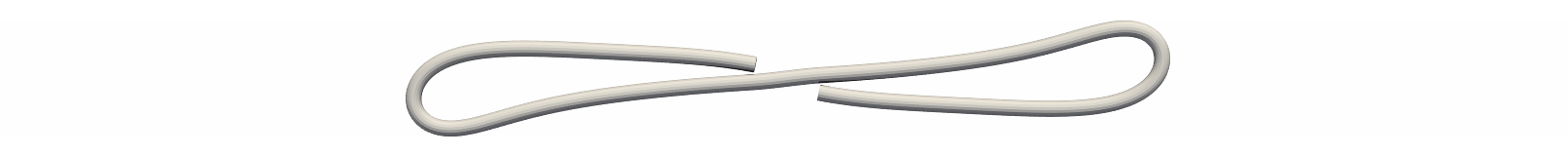} \\
                T = 25.5 & \includegraphics[width=0.8\textwidth]{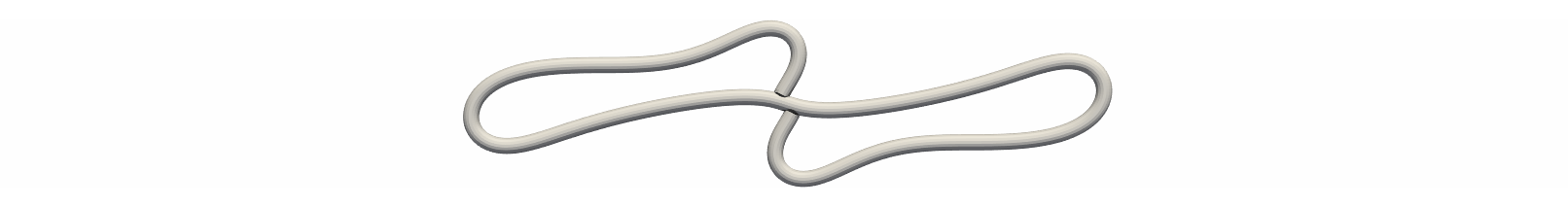} \\
                T = 28 & \includegraphics[width=0.8\textwidth]{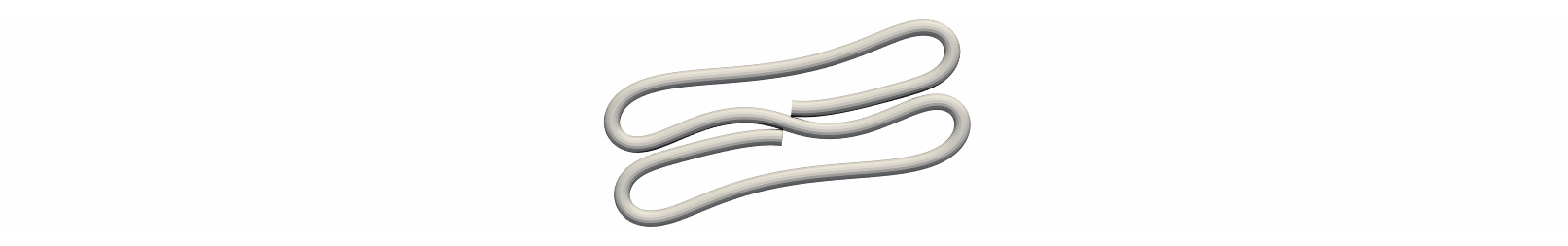} \\
                T = 30 & \includegraphics[width=0.8\textwidth]{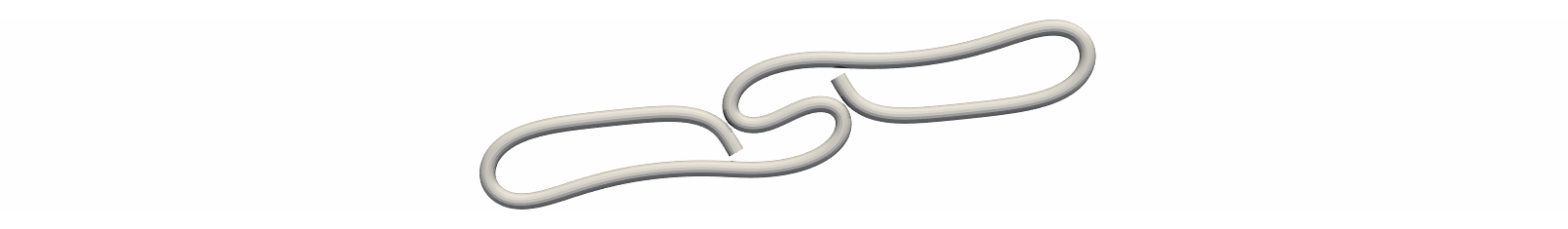} \\
                T = 30.75 & \includegraphics[width=0.8\textwidth]{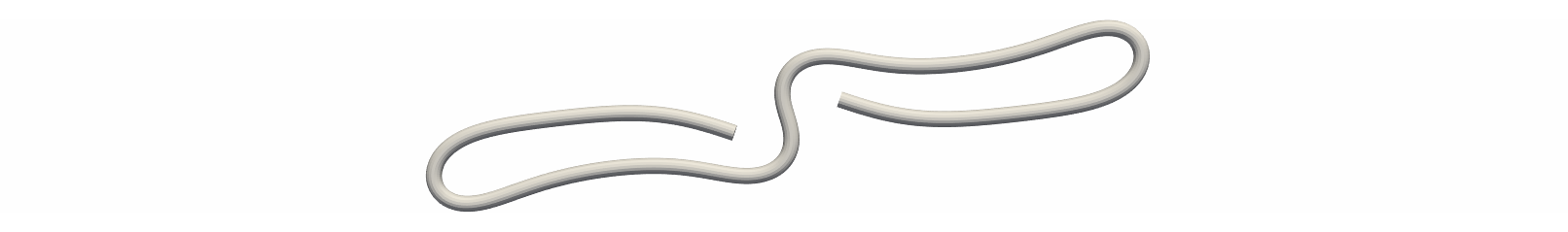} \\
                T = 31.25 & \includegraphics[width=0.8\textwidth]{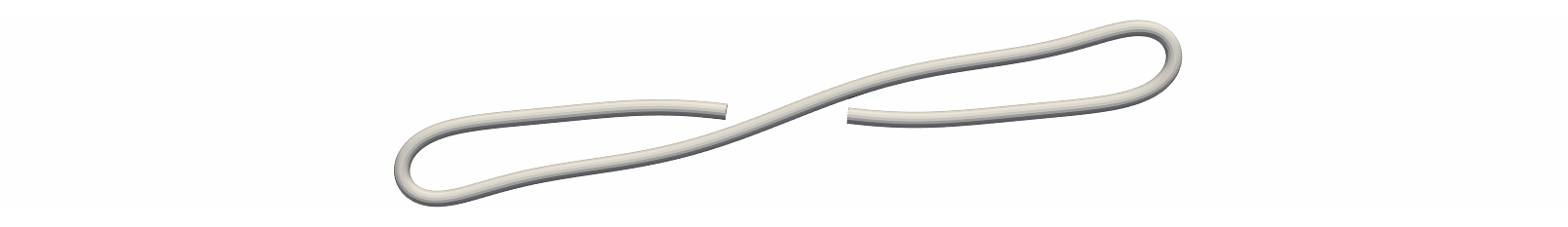} \\
        \end{tabular}
    \caption{Shape deformations of a fiber of length $L=0.931$ with corresponding elasto-viscous number 
$\bar{\mu}=7.51\times 10^{5}$.   }
    \label{long}
\end{figure}

\begin{figure}
        \centering
        \begin{tabular}{cc}
                T = 0 & \includegraphics[width=0.9\textwidth]{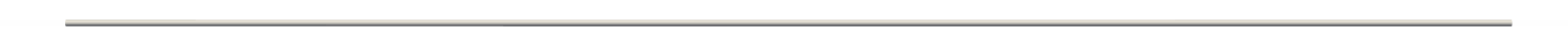} \\
                T = 10 & \includegraphics[width=0.9\textwidth]{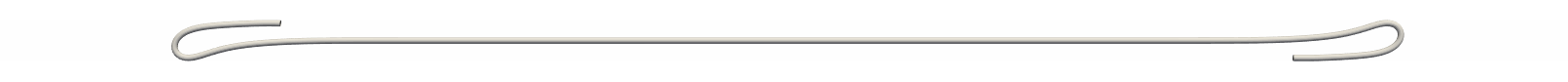} \\
                T = 24 & \includegraphics[width=0.9\textwidth]{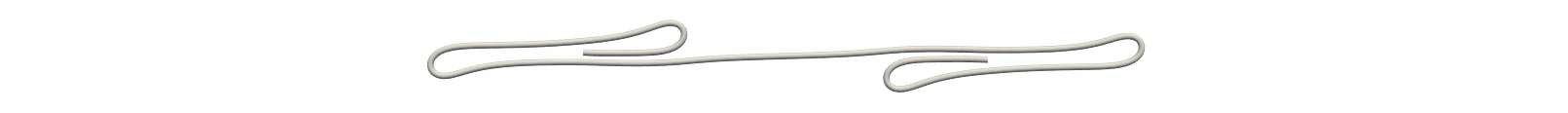} \\
                T = 31 & \includegraphics[width=0.9\textwidth]{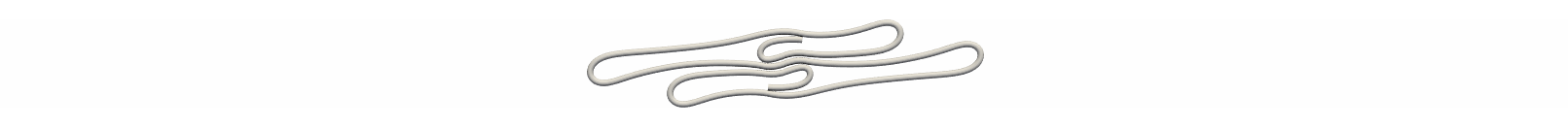} \\
                T = 36 & \includegraphics[width=0.9\textwidth]{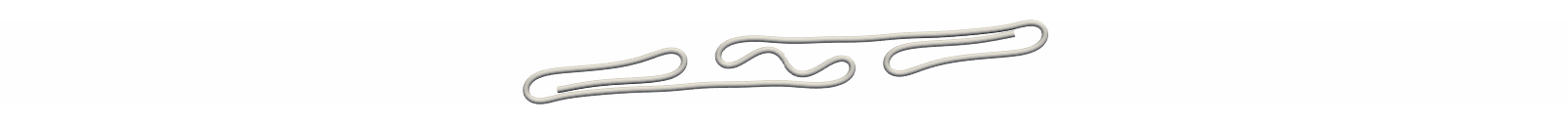} \\
                T = 39 & \includegraphics[width=0.9\textwidth]{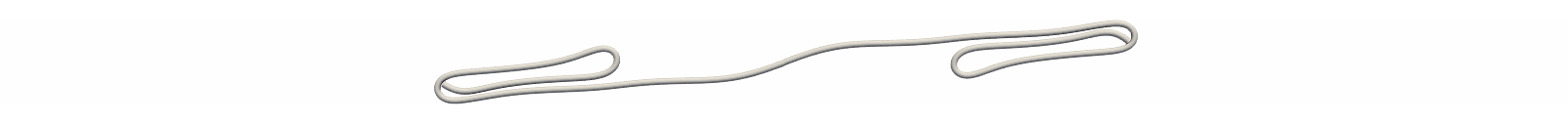} \\
                T = 44 & \includegraphics[width=0.9\textwidth]{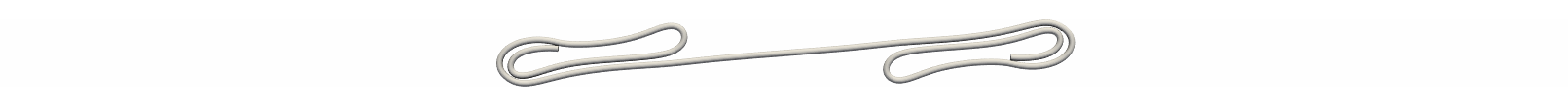} \\
                T = 50 & \includegraphics[width=0.9\textwidth]{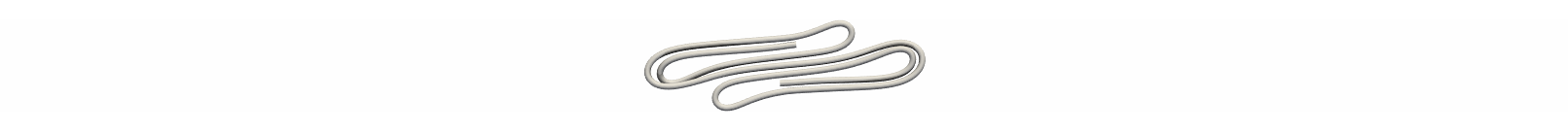} \\
                T = 52.5 & \includegraphics[width=0.9\textwidth]{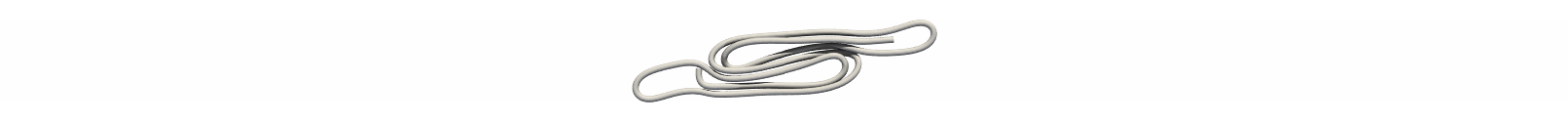} \\
                T = 55 & \includegraphics[width=0.9\textwidth]{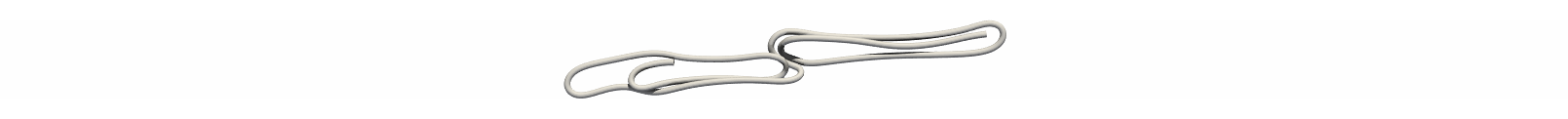} \\
                T = 60 & \includegraphics[width=0.9\textwidth]{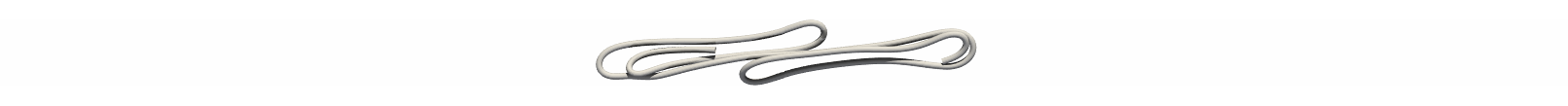} \\
                T = 70 & \includegraphics[width=0.9\textwidth]{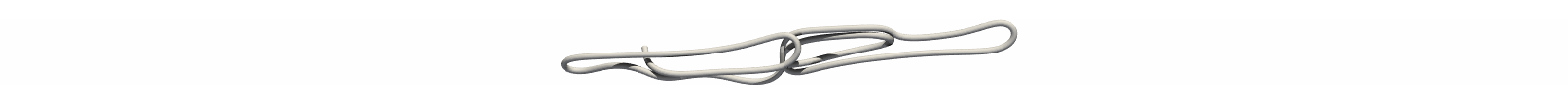} \\
                T = 85 & \includegraphics[width=0.9\textwidth]{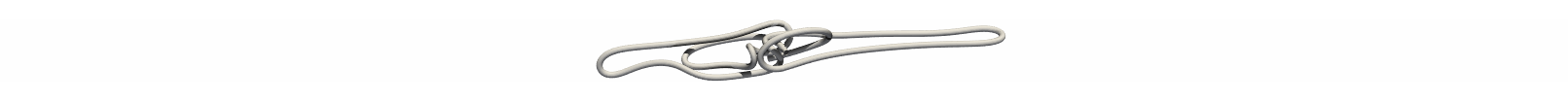} \\
                T = 110 & \includegraphics[width=0.9\textwidth]{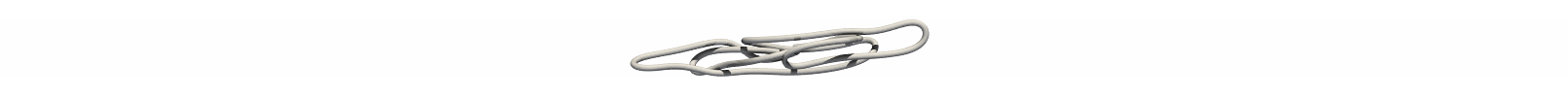} \\
        \end{tabular}
        \caption{Shape deformations of a fiber of length $L=2.28$ with corresponding elasto-viscous number 
$\bar{\mu} = 2.37\times 10^{7}$.  Note that after $T = 50$ the fiber centerline is no longer planar.  Numerical
fluctuations allow the fiber to exhibit a series of 3D coiled and entangled states.   }
    \label{longest}
\end{figure}

\begin{figure}
        \centering
        \includegraphics[width=0.8\textwidth]{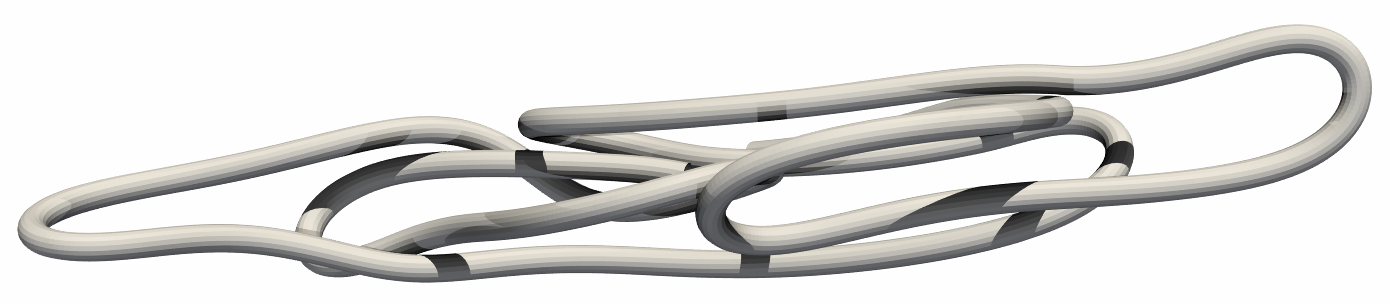}
        \caption{A zoomed-in image at $T = 110$ of the 3D entangled state of the fiber of length $L=2.28$ 
with corresponding visco-elastic number $\bar{\mu} = 2.37\times 10^{7}$.} 
        \label{3dblowup}
\end{figure}

We now examine the dynamics of longer fibers with values of elasto-viscous number 
$\bar{\mu}$ that transition from the snaking behavior shown in the third column of Figure \ref {tumble} to more complex
dynamics and shape evolution.  
Figure \ref{long} shows some time snapshots of a fiber of length $L = 0.931$ in shear with a corresponding
elasto-viscous number ($\bar{\mu}=7.51\times 10^{5}$) that is beyond those considered in \cite{liu2018morphological}.  
At time $T = 4.95$ we observe the emergence of the hooks at each end that are evident in snaking behavior.  However, it is worthwhile
to compare the snapshot of the $L=0.523$ fiber in Figure \ref {tumble} at $T=10.5$ to the longer fiber in
Figure \ref {long} at $T = 16.25$.  In the longer fiber, we see that additional hooks have formed at the ends, giving 
multiple local extrema in 
curvature along the fiber. The evolution of the curvature is shown in Figure \ref{curvature} (d), where we 
can see the propagation of traveling waves of curvature.  We remark that this fiber never did regain its 
straight shape.  Figure \ref {longest} shows the dynamics of an even longer fiber of length $L = 2.28$ that
exhibits even richer shape dynamics.  Again, the hooks appear at the fiber ends at $T = 10$, but the fiber is
long enough to support multiple coils as time evolves, as well as additional buckling sites in the middle.
Figure \ref {curvature} (e) shows the evolution of the curvature along this fiber with a cross-section at
time $T = 36$ indicated.  Figure \ref {curvature} (f) shows the curvature plotted at $T = 36$ as a function of 
arclength.  The multiple peaks in curvature can be identified in the fiber configuration also shown.  
In Figure \ref {longest} we can observe 
that at about $T=50$, the perturbations due to numerical fluctuations cause the centerline of
the fiber to move out of the plane, giving rise to the entanglement seen by Forgacs and Mason \cite {Forgacs1} and in \cite {yananthesis}. 
Figure \ref {3dblowup} shows a zoomed-in image of this
entanglement at time $T = 110$. 

%Note that there is no self-intersection of the fiber in this entangled state. 
%Because of finite time steps and finite regularization parameters, we cannot rule out self-intersection.
%While we have not implemented any repulsive forces here as fiber points get very close, such adjustments to the
%algorithm can be included.   

\begin{figure}
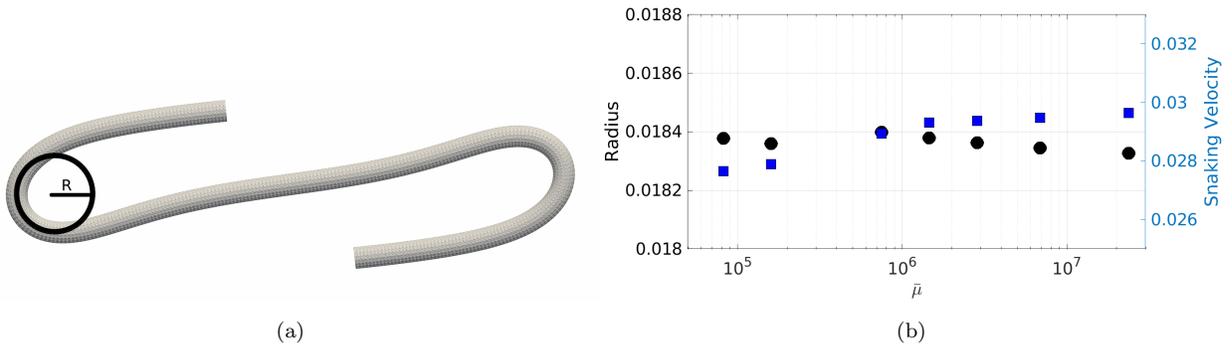

        \centering
        \subfloat[ ]{\includegraphics[width=0.5\textwidth]{{{radius_fiber}}}}\hfill
        \subfloat[  ]{\includegraphics[width=0.5\textwidth]{{{radius_snaking_mub}}}}\hfill
        \caption{(a) Radius of $J$-turn illustrated on a fiber. 
(b) For elasto-viscous numbers that exhibit $J$-turn formations initially, the computed radius of the hook and
the snaking velocity during the initial snaking.  Note that all simulations, independent of fiber length (and, hence,
visco-elastic number), result in radii and velocities that are nearly constant.}   
\label{jturn}
\end{figure}

Both Harasim et. al. \cite{Harasim1} and Liu et. al \cite {liu2018morphological} presented theoretical
analysis of the dynamics of the $J$ shape at the fiber ends that evolve during a snaking orbit.  In particular, it was
assumed that the hook developed is well-approximated by a semicircle of a fixed radius.  For fibers in our simulations
that did cross the threshhold of $\bar{\mu}$ at which snaking occurs, we measured the radius of such a circle
using a least squares fit (see Figure \ref {jturn} (a) ).  Even for large values of $\bar{\mu}$, the initiation of
bending exhibits the emergence of hooks.  Figure \ref {jturn} (b) shows that the emergent radii is nearly independent
of the length of the fiber and, hence, independent of $\bar{\mu}$.  Because our computations are
nearly at the limit of long filaments, this agrees with the predictions of \cite {Harasim1}.  
During the initiation of the snaking in each of the simulations (even those that go on to complex shape deformations with
no periodic orbits), we can measure the speed of propagation of the maximal curvature along the fiber arclengths.
This is equivalent to the slope of the traveling waves in the curvature surface plots in Figure \ref {curvature} (c-e).
Although not immediately apparent, because the range of the spatial axis is the arclength of the fiber which is different
in each panel, these
slopes are nearly equal. 
As with the hook radii, we find that these ``snaking velocities" are also nearly independent of  $\bar{\mu}$ (see
Figure \ref {jturn} (b)).

\subsection {Computational considerations}

\begin{figure}
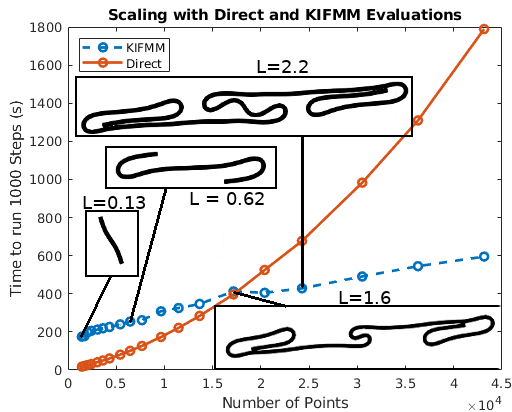

        \centering
     \includegraphics[width=0.45\textwidth]{{{scaling_with_inserts}}}
        \caption{Timings for regularized Stokeslet simulations
using direct $N^2$ force evaluation and kernel independent fast multipole (KIFMM) evaluation as a function of the number of discrete points
comprising the fiber surface.  Note that
the number of points increases as fiber length increases.  The insets show
a few fiber geometries at that value of $N$ surface points.}
       \label{scaling}
\end{figure}

To capture the complex coiling and entanglement dynamics of slender fibers, fine resolution of their
surface and along their length is required.  While the regularized Stokeslet formulation relies on fundamental
solutions of the Stokes equations and not a finite difference or finite element discretation of the surrounding 
three-dimensional fluid domain, a direct $N^2$ evaluation of $N$ velocities at $N$ nodes becomes prohibitive for
large $N$.  Figure \ref {scaling} presents the timings for 1000 time steps of the fiber-fluid system using 
direct summation and KIFMM summation. 
We see that for small fiber lengths, the overhead for KIFMM outweighs its benefits, but that a cross-over
occurs at about $N=1.7 \times 10^4$ nodes, corresponding to a fiber length of $L = 1.6$.  For the 
longest fiber of length $L = 2.2$ we already see a factor of 1.6 speed-up.  
This speed-up will be significant for simulations of multiple fiber dynamics. 

Note that in the simulations presented above, and in the entangled state shown in Figure \ref {3dblowup},
there is no self-intersection of the fiber. 
Because of finite time steps and finite regularization parameters, we cannot rule out self-intersection.
While we have not implemented any repulsive forces as fiber points get very close, such adjustments to the
algorithm can be included.

\section{Conclusions}

In summary, we have presented a computational method that captures complex shape deformations of long flexible fibers
in shear.  Using the method of regularized Stokeslets in conjunction with a kernel independent fast multipole method
we are able to simulate the dynamics of fibers at elasto-viscous numbers that are a couple of 
orders of magnitudes larger than those
reported using  slender body formulations.  
Because the fiber is represented by a spring-node system with individual spring elements, it is straightforward to
model fibers with inhomogenous material properties by altering connectivities and stiffness constants of the individual
elements.  Moreover, the implementation of KIFMM will allow us to probe the interaction of multiple fibers in
an array of background flows beyond a simple linear shear.

\section{Acknowledgements}
This research was supported by 
the National Science Foundation grant DMS-1043626 and the Gulf of Mexico Research Initiative.
The authors would like to thank Olivia du Roure, Yanan Liu, Anke Lindner and Michael Shelley for helpful discussions.

\bibliographystyle{plain}
\bibliography{shear}

\begin{thebibliography}{10}

\bibitem{cortez2005method}
R.~Cortez, L.~Fauci, and A.~Medovikov.
\newblock The method of regularized stokeslets in three dimensions: analysis,
  validation, and application to helical swimming.
\newblock {\em Physics of Fluids}, 17(3):031504, 2005.

\bibitem{cortez2001}
Ricardo Cortez.
\newblock The method of regularized stokeslets.
\newblock {\em SIAM Journal on Scientific Computing}, 23(4):1204--1225, 2001.

\bibitem{annrev2019}
Olivia du~Roure, Anke Lindner, Ehssan Nazockdast, and Michael Shelley.
\newblock Dynamics of flexible fibers in viscous flows and fluids.
\newblock {\em Ann. Rev. Fluid Mech.}, 51:539--572, 2019.

\bibitem{fd2006}
L.~Fauci and R.~Dillon.
\newblock Biofluidmechanics of reproduction.
\newblock {\em Annu. Rev. Fluid. Mech.}, 38:371--394, 2006.

\bibitem{Flores2005}
H.~Flores, E.~Lobaton, S.~M{\'e}ndez-Diez, S.~Tlupova, and R.~Cortez.
\newblock A study of bacterial flagellar bundling.
\newblock {\em Bulletin of Mathematical Biology}, 67(1):137--168, 2005.

\bibitem{Forgacs1}
O.L. Forgacs and S.G. Mason.
\newblock Particle motions in sheared suspensions: X {O}rbits of flexible
  threadlike particles.
\newblock {\em J. Colloid Sci.}, 14:473--491, 1959.

\bibitem{Harasim1}
M.~Harasim, B.~Wunderlich, O.~Peleg, M.~Kroger, and A.~Bausch.
\newblock Direct observation of the dynamics of semiflexible polymers in shear
  flow.
\newblock {\em Phys. Rev. Lett.}, 110:108302, 2013.

\bibitem{Kantsler1}
V.~Kantsler and R.~Goldstein.
\newblock {Fluctuations, dynamics, and the stretch-coil transition of a single
  actin filament in extensional flow}.
\newblock {\em Phys. Rev. Lett.}, 108:038103, 2012.

\bibitem{karp1998motion}
Lee Karp-Boss and Peter~A Jumars.
\newblock Motion of diatom chains in steady shear flow.
\newblock {\em Limnol. Oceanogr.}, 43(8), 1998.

\bibitem{lagrone1}
John Lagrone, Ricardo Cortez, and Lisa Fauci.
\newblock Elastohydrodynamics of swimming helices: effects of flexibility and
  confinement.
\newblock {\em Phys. Rev. Fluids}, 2019.

\bibitem{peskinlim2004}
S.~Lim and C.S. Peskin.
\newblock Simulations of the whirling instability by the immersed boundary
  method.
\newblock {\em SIAM J. Sci. Comp.}, 25(6):2066--2083, 2004.

\bibitem{yananthesis}
Yanan Liu.
\newblock {\em Dynamics of flexible and Brownian filaments in viscous flows}.
\newblock PhD thesis, l'Universite Paris Diderot, 2018.

\bibitem{liu2018morphological}
Yanan Liu, Brato Chakrabarti, David Saintillan, Anke Lindner, and Olivia
  du~Roure.
\newblock Morphological transitions of elastic filaments in shear flow.
\newblock {\em Proceedings of the National Academy of Sciences},
  115(38):9438--9443, 2018.

\bibitem{malhotra_pvfmm_2015}
Dhairya Malhotra and George Biros.
\newblock {{PVFMM}}: {{A Parallel Kernel Independent FMM}} for {{Particle}} and
  {{Volume Potentials}}.
\newblock {\em Communications in Computational Physics}, 18(03):808--830, 2015.

\bibitem{Manikantan1}
Harishankar Manikantan and David Saintillan.
\newblock {Buckling transition of a semiflexible filament in extensional flow}.
\newblock {\em Phys.Rev. Lett.}, 92, 2015.

\bibitem{Nguyen2017}
F.~Nguyen and M.~Graham.
\newblock Buckling instabilities and complex trajectories in a simple model of
  uniflagellar bacteria.
\newblock {\em Biophysical Journal}, 112(5):1010--1022, 2017.

\bibitem{Fauci4}
H.~Nguyen and L.~Fauci.
\newblock Hydrodynamics of diatom chains and semiflexible fibres.
\newblock {\em J. Royal Soc. Interface}, 11:20140314, 2014.

\bibitem{jedd2015}
L.~Pieuchot, J.~Lai, R.~Loh, F.~Leong, K-H Chiam, J.~Stajich, and G.~Jedd.
\newblock Cellular subcompartments through cytoplasmic streaming.
\newblock {\em Devel. Cell}, 34:410--420, 2015.

\bibitem{Pozrikidis:92}
C.~Pozrikidis.
\newblock {\em Boundary integral and singularity methods for linearized viscous
  flow}.
\newblock Cambridge texts in applied mathematics. Cambridge University Press,
  1992.

\bibitem{Quennouz1}
N.~Quennouz, M.~Shelley, O.~du~Roure, and A.~Lindner.
\newblock Transport and buckling dynamics of an elastic fibre in a viscous
  cellular flow.
\newblock {\em J. Fluid Mech.}, 769:387--402, 2015.

\bibitem{shelley2016}
Michael Shelley.
\newblock The dynamics of microtubule/motor-protein assemblies in biology and
  physics.
\newblock {\em Ann. Rev. Fluid Mech.}, 48:487--506, 2016.

\bibitem{ts2007}
A.K. Tornberg and M~Shelley.
\newblock Simulating the dynamics and interactions of elastic filaments in
  {S}tokes flow.
\newblock {\em J. Comp. Phys.}, 196:8--40, 2007.

\bibitem{Wandersman1}
E.~Wandersman, N.~Quennouz, M.~Fermigier, A.~Lindner, and O.~du~Roure.
\newblock Buckled in translation.
\newblock {\em Soft Matter}, 6:5715--5719, 2010.

\bibitem{yan_flexibly_2018}
Wen Yan and Michael Shelley.
\newblock Flexibly imposing periodicity in kernel independent {{FMM}}: {{A}}
  multipole-to-local operator approach.
\newblock {\em Journal of Computational Physics}, 355(Supplement C):214--232,
  2018.

\bibitem{qiang2017}
Q.~Yang and L.~Fauci.
\newblock Dynamics of a macroscopic elastic fibre in a polymeric cellular flow.
\newblock {\em J. Fluid Mech.}, 817:388--405, 2017.

\bibitem{ying_kernel-independent_2004}
Lexing Ying, George Biros, and Denis Zorin.
\newblock A kernel-independent adaptive fast multipole algorithm in two and
  three dimensions.
\newblock {\em Journal of Computational Physics}, 196(2):591--626, 2004.

\end{thebibliography}

\end{document}